\documentclass{article}

\usepackage[margin=1in, paperwidth=8.5in, paperheight=11in]{geometry}

\usepackage{cite}
\usepackage{amsmath}
\usepackage{graphicx}
\usepackage{subfigure}
\usepackage{verbatim}
\usepackage{amsfonts}
\usepackage{booktabs}
\usepackage{pstricks}

\usepackage{multirow}

\usepackage{color}
\usepackage{soul}

\xdefinecolor{lightgray}{rgb}{.88,.88,.88}

\title{Sensitivity of Double-Sided Split Ring Resonator Arrays to Fabrication Tolerances}
\author{Frank Trang, Edward F. Kuester, Horst Rogalla, and Zoya Popovi\'{c}\\%
Department of Electrical, Computer, and Energy Engineering\\%
 University of Colorado, Boulder, CO  80309
}
\date{}

\begin{document}

\maketitle

\begin{abstract}
We present a study of the effects of fabrication tolerances on the performance of a planar array of double-sided split-ring resonators, printed on two sides of a dielectric substrate and fabricated using a printed circuit board (PCB) milling machine. The array is simulated and measured in an X-band waveguide, and the measured resonant frequency is found to be 6.3\% higher than the predicted one. The sensitivity of the frequency response to several possible fabrication and measurement errors is investigated, and the dominant effect identified and demonstrated experimentally and in simulation.

\end{abstract}

\section{Introduction}
\label{sec:introduction}

Split ring resonators, first introduced by Pendry \textit{et al.}~\cite{798002}, have been a popular choice as building blocks for many metamaterial designs, including double negative index materials \cite{Shelby06042001}, electromagnetic cloaks \cite{Schurig10112006}, etc.  Many of the characteristic properties of metamaterials occur at and close to the resonances of the constituent unit-cell scatterers.  However, for resonant structures, imperfections due to the fabrication process can greatly affect the performance.

This paper presents experimental and simulated results for the frequency response of a double-sided split ring resonator (DSRR) array designed to operate at X-band and measured inside a waveguide, as shown in Figure \ref{fig:dsrr_in_waveguide}.  Our main purpose is to investigate the impact fabrication tolerances have on the resonances of the DSRR array, with the following imperfections studied in detail:  tilt of substrate when placed inside the waveguide, imperfect overlap of the two SRRs on the two sides of the substrate, air gaps along the waveguide walls, varying line width of the copper strips, varying substrate thickness, permittivity inhomogeneity, and grooves in the substrate adjacent to the metal traces.  Simulated results of the DSRR array in an ideal case are compared with those from measurements, which show that relatively minor errors and fabrication tolerances, which are sometimes overlooked, can have a large effect on the resonant frequency.

\section{DSRR Design and Fabrication}
\label{sec:dsrr_design_and_fabrication}

Typically, arrays of split rings are designed for free-space plane-wave filtering and related functions.  Free-space measurements require large arrays and are difficult to calibrate.  In this work, we characterize small arrays of split-ring resonators in a waveguide environment, which can be both easily simulated and calibrated.  After good agreement is obtained between simulations and measurements, free-space simulations can be trusted for design.

\begin{figure}
   \centering

   \subfigure[]
   {
      \label{fig:dsrr_in_waveguide}
      \includegraphics{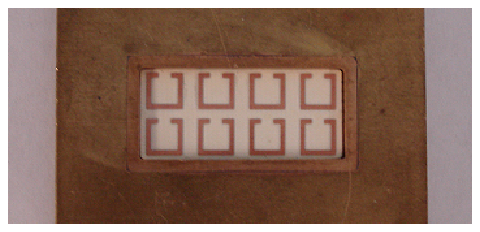}
   }
   \subfigure[]
   {
      \label{fig:dsrr_unit_cell}
      \includegraphics{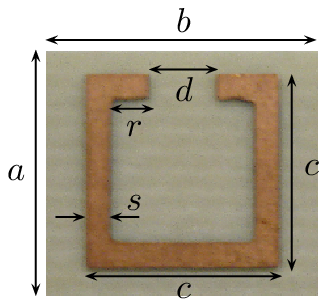}
   }
   
   \subfigure[]
   {
      \label{fig:dsrr_side_view}
      \includegraphics{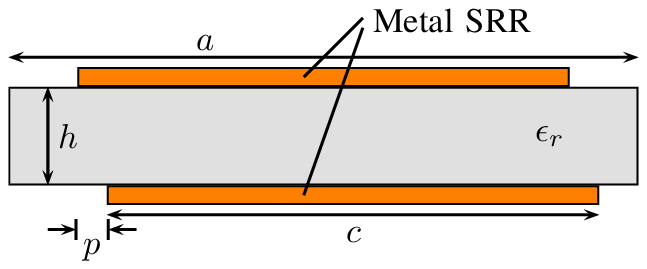}
   }

   \caption{(a) Photograph of 8-element DSRR placed in a transverse plane of a WR-90 TE$_{10}$ waveguide.  (b)  Unit cell of the DSRR.  (c)  Side view of a unit cell (not to scale).}
   \label{fig:dsrr}
\end{figure}

The unit cell in Figure \ref{fig:dsrr_unit_cell} is designed with dimensions that result in a resonance at 9.2\,GHz in transmission.  A double-sided structure is chosen to provide symmetry.  The dimensions of a DSRR unit cell are labeled in Figure \ref{fig:dsrr_unit_cell}, where $a$=5.08\,mm, $b$=5.715\,mm, $c$=4\,mm, $d$=1.4\,mm, $r$=0.8\,mm, $s$=0.5\,mm.   The DSRRs are made up of 35\,$\mu$m thick copper rings printed on the two opposite sides of a $h$=762\,$\mu$m thick slab of Rogers 4350B substrate, which has a nominal relative permittivity $\epsilon_r$=3.66, Figure \ref{fig:dsrr_side_view}.  Five 2x4 arrays of DSRRs, referred to as DSRR1$-$DSRR5, each consisting of eight unit cells with the vertical periodicity of 5.08\,mm and lateral periodicity of 5.715\,mm, were fabricated for measuring in an X-band waveguide.  The outer dimensions of the unit cell were chosen to allow an integer number of the DSRRs to fit inside a standard X-band waveguide.  An LPKF ProtoMat S62 printed circuit board (PCB) milling machine was used to define the metal rings and the alignment between the two sides was done with 1.1mm diameter alignment holes in each corner.

\section{Measurements and Simulations}
\label{sec:measurements_and_simulations}

\begin{figure}
   \centering
   \subfigure[]
   {
      \includegraphics{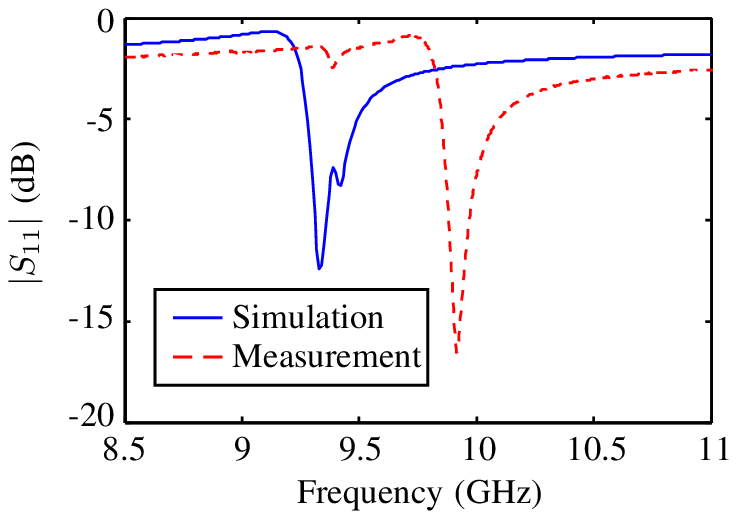}
      \label{fig:measurement_vs_simulation_s11}
   }
   \subfigure[]
   {
      \includegraphics{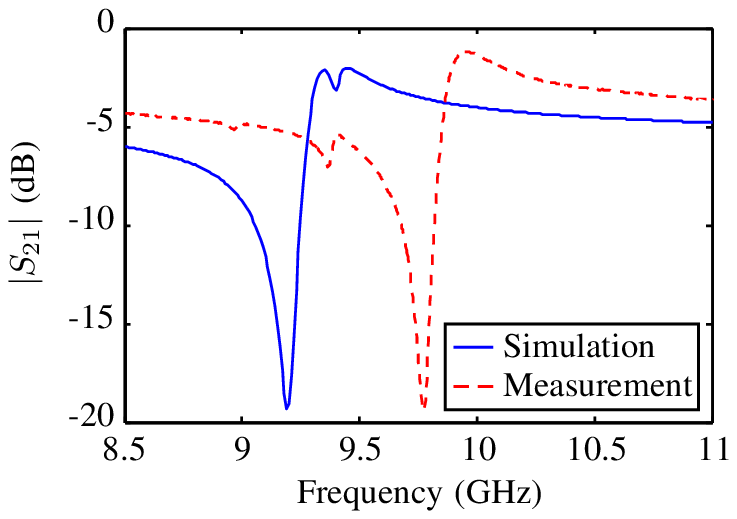}
      \label{fig:measurement_vs_simulation_s21}
   }
   \caption{Simulated and measured reflection (a) and transmission (b) coefficient magnitudes of the DSRR1 array inside the waveguide.}
   \label{fig:sim_meas_s_parameters}
\end{figure}

The DSRR structure placed inside a WR-90 waveguide was simulated in Ansoft HFSS, a finite element method (FEM) solver, with the simulated reflection ($S_{11}$) and transmission ($S_{21}$) coefficient magnitudes shown together with the measurements in Figure \ref{fig:sim_meas_s_parameters}.  The Agilent E8364B PNA was calibrated from 8.2$-$12.4\,GHz using the WR-90 Maury Microwave waveguide calibration standards.

\begin{figure}[h]
   \centering
   \includegraphics{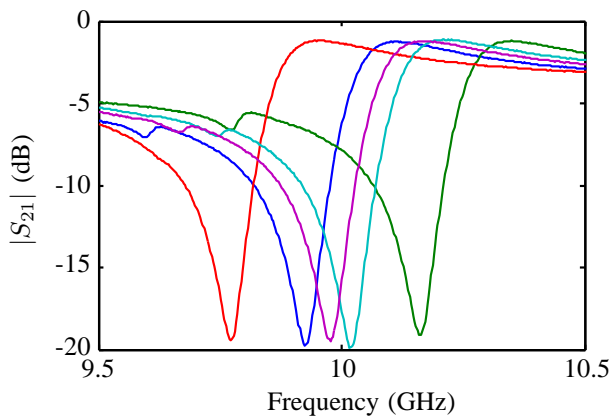}
   \caption{Measured transmission coefficients for DSRR1$-$DSRR5 (with minima ordered left to right).}
   \label{fig:overlayed_measured_s21}
\end{figure}

Figure \ref{fig:overlayed_measured_s21} shows the measured $|S_{21}|$ for each of the five arrays, showing a 380\,MHz variation which can only be attributed to variations in fabrication.  Rather surprisingly, we observed a 580\,MHz, or 6.3\%, upward shift in resonant frequency of the transmission coefficient relative to the simulation for DSRR1.  This discrepancy is not expected and can be due to either simulation, measurement, or fabrication errors.  FEM (HFSS) is well suited for closed structures and our experience is that we can trust the simulation.  The excitation for the simulation were waveports and the meshing was varied to check for convergence.

\section{Tolerance Studies}
\label{sec:tolerance_studies}

\begin{figure}[h]
   \centering
   \includegraphics{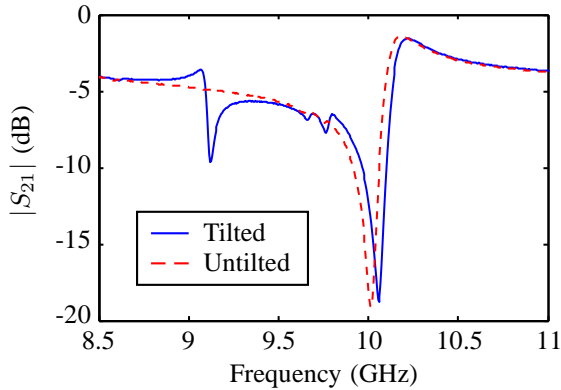}
   \caption{A comparison of the transmission coefficients for the cases of a tilted (10 degree) and an untilted substrate in a waveguide for DSRR4.}
   \label{fig:tilt_vs_untilt_s21}
\end{figure}

The two practical sources of errors can be divided into measurement errors (calibration and misalignment) and fabrication errors.  To rule out calibration errors, two separate calibration methods were employed, a standard Thru-Reflect-Line (TRL) method and a Short-Short-Load-Thru (SSLT) method using Maury X7005S calibration standards.  The results, not shown, match well, thus ruling out calibration errors.  Another measurement error is a possible tilt of the structure inside the waveguide.  With a large tilt of 10$^\circ$ introduced in the measurement, there is a slight shift (0.4\%) in the resonance, as shown in Figure \ref{fig:tilt_vs_untilt_s21}.  However, it does not explain the major shift in the main resonance.

Several obvious fabrication errors include: (1) imperfectly overlapping rings, described by $p$ in Figure \ref{fig:dsrr_side_view}; (2) an air gap along the waveguide walls; (3) varying line width of the copper strip; (4) varying substrate thickness; (5) permittivity inhomogeneity.  For the fabrication techniques used in this work, typical deviations were taken into account to simulate their effects on the resonant frequency, as summarized in Table~\ref{table:tolerance_study}.  The following conclusions can be made:

\begin{itemize}
   \item
      Misalignment between the front and back side split-rings, described by the overlap $p$, does not cause significant shift in the resonant frequency.

   \item
      The 0.5-mm line width can vary if the milling bit is dulled after extended use.  A 50\,$\mu$m increase in width shifts the resonance upward by 0.11\%, while a 50\,$\mu$m decrease shifts it downward by 1.1\%.

   \item
      For a large sheet of substrate, there will likely be areas where the sheet is thinner or thicker than the manufacturer's specified value.  Thus, the thickness of the substrate was varied by +18\,$\mu$m and -12\,$\mu$m in the simulations.  The tabulated results in Table \ref{table:tolerance_study} show the resonance was effected by at most 0.5\%.

   \item
      The permittivity has been shown to vary across a substrate sheet, e.g. \cite{4136048},\cite{99776}.  The nominal permittivity value quoted by the manufacturer was 3.66.  In simulation, this was varied between 3.5 and 3.8, which resulted in a total resonance shift of 1.5\%.  Table~\ref{table:tolerance_study} shows the effects of this deviation.  For further verification, a test structure of microstrip lines was milled on the same sheet of substrate used for the DSRR array.  Using an extraction method discussed by Das \textit{et al.} \cite{1133722}, the effective dielectric constant was then measured and a value of $\epsilon_r$=3.65 was extracted at 9\,GHz.

\end{itemize}

\begin{table}
   \caption{Fabrication imperfections and their effects on the resonant frequency. (V$-$vertical offset, H$-$horizontal offset)}
   \label{table:tolerance_study}
   \vspace{.25\baselineskip}
   
   \centering
   \begin{tabular}{ c c c c }
      \toprule
      Parameters & Value & $f_r$ (GHz) & \% Shift\\
      \midrule
      \multirow{4}{*}{Overlap Offset $p$ ($\mu$m)} & 40 (V) & 9.22 & +0.33\\
      & 80 (V) & 9.14 & -0.54\\
      & 50 (H) & 9.18 & -0.11\\
      & 100 (H) & 9.07 & -1.31\\
      \midrule
      \multirow{2}{*}{Line Width $s$ (mm)} & 0.45 & 9.2 & +0.11\\
      & 0.55 & 9.09 & -1.09\\
      \midrule
      \multirow{2}{*}{Substrate Thickness $h$ ($\mu$m)} & 750 & 9.19 & 0\\
      & 780 & 9.14 & -0.54\\
      \midrule
      \multirow{2}{*}{Relative Permittivity $\epsilon_r$} & 3.5 & 9.2 & +0.11\\
      & 3.8 & 9.06 & -1.41\\
      \bottomrule
   \end{tabular}
\end{table}

An additional possible shift could be attributed to air gaps between the substrate and waveguide walls.  Air gaps along the top and bottom wall dominate since the waveguide is excited with a TE$_{10}$ wave mode, for which the electric field goes to zero on the side walls.  When an air gap as large as 150\,$\mu$m is introduced in simulation, the resonance shifts by only 0.65\%.

\begin{figure}
   \centering
   \subfigure[]{\includegraphics{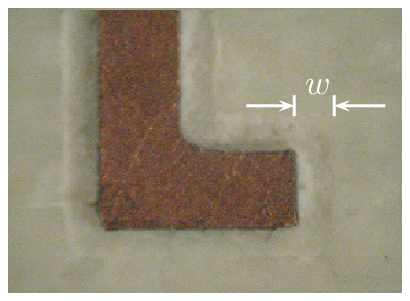}}
   \subfigure[]{\includegraphics{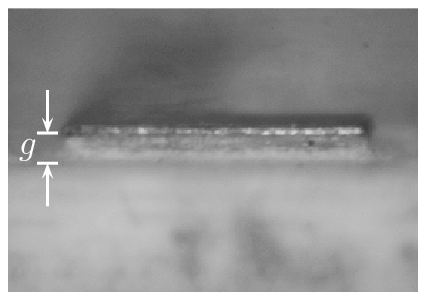}}
   
   \subfigure[]{\includegraphics{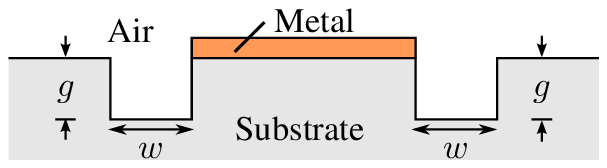}}
   \caption{Microscope photographs showing the width ($w \approx 254\,\mu$m) and depth ($g \approx 105\,\mu$m) of a groove from the top (a) and side (b).  (c) shows a sketch of a cross-section with relevant dimension parameters (not to scale).}
   \label{fig:substrate_grooves}
\end{figure}

In summary, none of the obvious fabrication or measurement errors account for more than 1.41\% of the resonance shift in the worst case, indicating that there is some discrepancy that we have not yet taken into account.  A closer examination of the fabricated DSRR structure shows that the milling machine bit leaves grooves in the dielectric adjacent to the metal trace, as shown in Figure 4.  There has been some previous work \cite{4545894, MOP:MOP25724, 1017891} reporting a change in effective permittivity of planar transmission lines when grooves are present on the substrate.  However, the work presented in these papers is on guided wave (microstrip) structures and very different in nature from the work discussed in our paper.  It is not possible to carry over conclusions directly from a guided wave structure to a free-space structure; one involves a change in phase velocity, while the other is effectively a change in the resonant lumped element shunt loading of a waveguide.  Nevertheless, we examined a section of the SRR through a microscope, and observed grooves, as shown in Figure \ref{fig:substrate_grooves}.  The grooves around the copper ring edges have a width approximately equal to the width of the milling bit (254\,$\mu$m) and a depth approximately three times the thickness of the copper (3$\times$35$\,\mu$m).  We expected that these minor imperfections might play some role in altering the reflection and transmission responses of the DSRR array, but as shown below, they turned out to be the main factor in the shifting of the resonances.  A nice PCB fabrication guideline by Trescases \cite{pcb_fab_guideline} discusses the procedure for adjusting the proper depth of the milling bit, thus reducing the depth of the grooves.

\begin{figure}[h]
   \centering
   \includegraphics{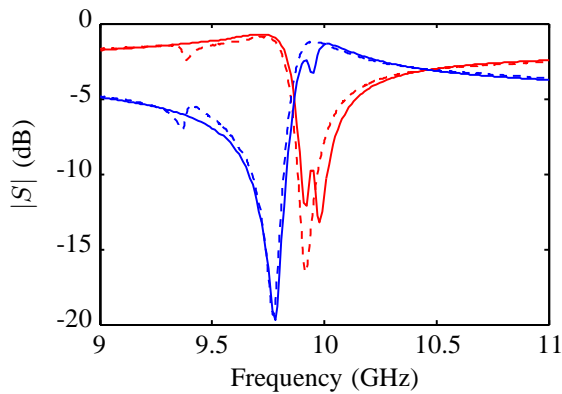}
   \caption{The measured transmission (dashed blue) and reflection (dashed red) coefficients shown together with the simulation (solid lines) for DSRR1, which includes identical grooves on both sides with a groove width $w$=254\,$\mu$m and depth $g$=107\,$\mu$m.  The ring dimensions are as specified in section \ref{sec:dsrr_design_and_fabrication} with $a$=5.08\,mm, $b$=5.715\,mm, $c$=4\,mm, $d$=1.4\,mm, $r$=0.8\,mm, $s$=0.5\,mm.}
   \label{fig:groove_overlay_plot}
\end{figure}

The nominal (HFSS) design was modified to include a groove width of 254\,$\mu$m and a depth of 107\,$\mu$m gives the best agreement with the experimental results.  The new scattering parameter are compared to the measurement in Figure \ref{fig:groove_overlay_plot}.  The two curves overlap closely.  We can therefore conclude that the 6.3\% shift of the resonant frequency is caused primarily by the presence of the grooves.

We further studied the frequency responses resulting from varying the groove width and depth, with the results summarized in Table \ref{table:groove_table}.  The values chosen are comparable to the size of milling bits commonly available.  The results show that varying either the width or the depth of grooves shifts the resonant frequency to higher values.

\begin{table}
   \caption{Resonant frequencies with varied groove dimensions}
   \label{table:groove_table}
   \vspace{.25\baselineskip}
   \centering
   \begin{tabular}{ccc}
      \toprule
      Depth $g$ ($\mu$m) & Width $w$ ($\mu$m) & $f_r$ (GHz) \\
      \midrule
      50.8 & 127 & 9.48 \\
      50.8 & 254 & 9.54 \\
      76.2 & 127 & 9.58 \\
      76.2 & 254 & 9.66 \\
      107.0 & 254 & 9.78 \\
      \bottomrule
   \end{tabular}
\end{table}

\begin{figure}
   \centering
   \subfigure[]{
      \includegraphics{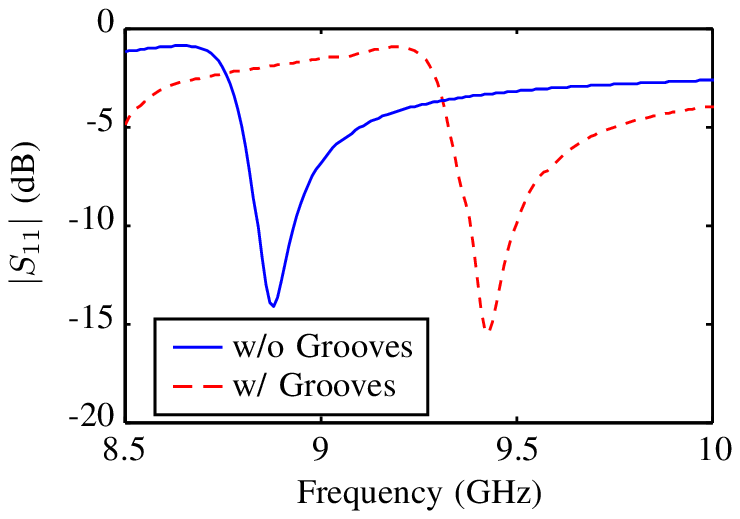}
      \label{fig:s11_wo_w_grooves}
   }
   \subfigure[]{
      \includegraphics{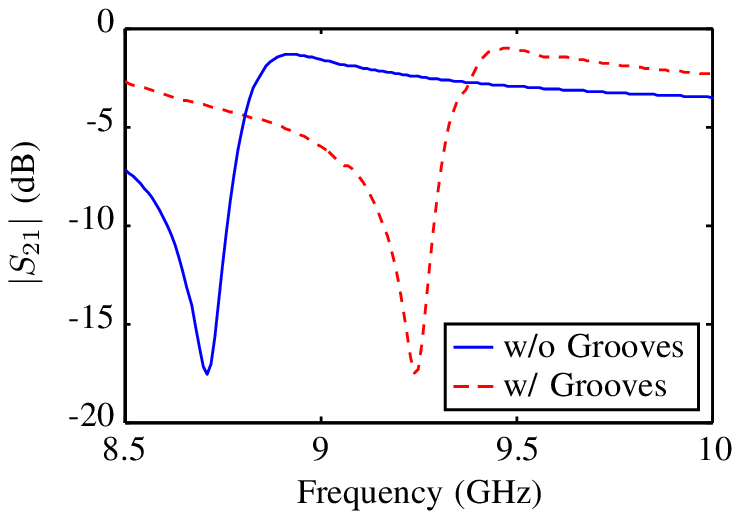}
      \label{fig:s21_wo_w_grooves}
   }
   \caption{Simulated reflection (a) and transmission (b) coefficient magnitudes of a single-sided SRR array with and without the grooves adjacent to the copper traces inside the waveguide.}
   \label{fig:single_sided_srr_s_parameters}
\end{figure}

A single-sided SRR structure with the same dimensions as those of the DSRRs specified in section \ref{sec:dsrr_design_and_fabrication} was also studied through simulations.  We expect the grooves in this structure to shift the resonance higher in frequency.  First the structure without the grooves is simulated, with the results shown as solid blue lines in Figure \ref{fig:single_sided_srr_s_parameters}.  Next, grooves with width $w$=254\,$\mu$m and depth $g$=107\,$\mu$m were added adjacent to the copper traces, with the results shown as the dashed red lines in Figure \ref{fig:single_sided_srr_s_parameters}.  The new simulation shows 6.2\% and 6.1\% upward shifts in the reflection and transmission resonant frequency, respectively.  From the results, we can make a stronger case that the effect of the grooves is not confined only to the DSRR design, but can be seen in other resonant circuits.

\section{Conclusion}
\label{sec:conclusion}

We have discovered that grooves in the dielectric substrate caused by a milling procedure can result in large changes in the resonant frequency of a planar array of DSRRs.  Although some previous researchers have found small effects of grooves in the substrate surface, our work demonstrated that resonances can be dramatically altered by this seemingly minor deviation from the design.  This illustrates the importance of very precise modeling whenever such resonant behavior is present.

\bibliographystyle{IEEEtran}
\bibliography{IEEEabrv,references}

\end{document}